\documentclass[aps,pra,showpacs,twocolumn]{revtex4}

\usepackage{amssymb}
\usepackage{epsfig}
\usepackage{graphicx}
\usepackage{amsmath}
\usepackage{array,color}

\begin{document}

\title{Three dimensional ring vortex solitons and their stability in Bose-Einstein condensates under magnetic confinement}
\author{Ji Li$^1$, Deng-Shan Wang$^2$, Zhi-Yong Wu$^{1,3}$, Yan-Mei Yu$^{1,\footnote{ymyu@iphy.ac.cn}}$
, and Wu-Ming Liu$^1$ }
\address{$^1$Beijing National Laboratory for Condensed Matter Physics, Institute of Physics, Chinese Academy of Sciences, Beijing 100190, China}
\address{$^2$School of Science, Beijing Information Science and Technology University, Beijing 100192, China}
\address{$^3$Department of Applied Physics,Yanshan University,Qinhuangdao, 066004, China.}
\date{\today}
\begin{abstract}
The three-dimensional study of the ring vortex solitons is conducted for both attractive and repulsive BECs subject to harmonic potential confinement. A family of stationary ring vortex solitons, which is defined by the radial excitation number and the winding number of the intrinsic vorticity, are obtained numerically for a given atomic interaction strength. We find that stabilities of the ground and radially excited states of the ring vortex soliton are dependent on the winding number differently. The ground state of the ring vortex soliton with the large winding number is unstable dynamically against random perturbation. The radially excited state of the ring vortex soliton with large winding number corresponds to the increased collapse threshold, and therefore can be made stable for sufficiently small atomic interaction strengths. The ground and radially excited states also demonstrate different dynamical evolutions under large atomic interaction strengths. The former exhibits simultaneously symmetrical splitting in the transverse plane, while the latter displays periodically expand-merge cycles in the longitudinal direction.
\end{abstract}

\maketitle
\par
\section{Introduction}
The realization of Bose-Einstein condensates (BECs) in dilute quantum gases has drawn a great deal of interest in vortices \cite{Matthews-PRL-1999,Fetter-JPCM-2001}. Recently, considerable efforts are aimed at the prediction of settings supporting stable multidimensional solitons with intrinsic vorticity.  One such structure is the ring vortex soliton, a soliton loops back on itself to form a ring and the phase of the wave function winds through an integer multiple of $2\pi$ radians around the vortex line. Such ring vortex soliton can be defined by the quantum number set ($n,S$), where $n$ is the radial excitation number, equal to the number of the rings, and $S$ is the winding number of the intrinsic vorticity. One example of the ring vortex soliton, being called ``ring-profile solitary waves'', has been studied in the context of optics \cite{Firth-PRL-1997,Neshev-APBLO-1997}. Another example of the ring vortex soliton, called ``vortex tori", has also been predicted in three dimensions (3D) cubic-quintic Ginzburg-Landau equation \cite{Mihalache-PRL-2006,Mihalache-PRL-2010}. The ring dark soliton is first introduced in the repulsive BECs \cite{Theocharis-PRL-2003}. In the recent binary BEC experiment, the ring bright solitons are also observed \cite{Mertes-PRL-2007,Scherer-PRL-2010}, which is studied further in the theoretical works \cite{Law-PRL-2010}.

The most simple structure of the ring vortex soliton is of one ring in the radial direction, i.e., $n$=1, which can be regarded as the radial ground state, and here we call it as single ring vortex soliton (SRVS). The bright vortex soliton created in the attractive and dipolar BECs \cite{Ueda-PRL-2002,Ueda-PRL-2003,Ueda-PRA-2004,Mihalache-PRA-2006,Malomed-PLA-2007,Zaliznyak-PLA-2008} can be regarded as one typical example of SRVS. As we are known, the atomic interaction in BECs can be controlled by the Feshbach resonance technique \cite{Inouye-Nature-1998,Kevrekidis-PRL-2003,Pollack-PRL-2009} to change the strength and sign of the interaction. Therefore, great interests are motivated to investigate the properties of the vortex soliton under the nonlinearity of the atomic interaction. It is predicted that SRVS is stable under the 3D harmonic trapping potential for sufficient small interaction strength \cite{Ueda-PRL-2002,Ueda-PRL-2003,Pu-PRA-1999}, while for large interaction strength the dynamical instability phenomena, such as split-merge cycles of particles \cite{Ueda-PRL-2002,Mihalache-PRA-2006,Salasnich-PRA-2009} and intertwining of doubly quantized vortex \cite{Mottonen-PRA-2003,Mateo-PRL-2006,Huhtamaki-PRL-2006}, occur in the unstable regime.

An infinite sequence of radially excited stationary states of the ring vortex have been predicted to Gross-Pitaevskii equation (GPE) in two dimensions (2D) and sphere shell in three dimensions (3D) \cite{Carr-PRA-2006,Carr-PRL-2006}. The radial excited state is of multiple concentric density-wave rings, i.e., $n\geq2$, and here we call it as the multiple ring vortex soliton (MRVS). The one dimensional (1D) and 2D solutions of the MRVS are also proposed for a specific spatially-modulated nonlinearity \cite{Wu-PRA-2010,Wang-PRA-2011,Tian-PRA-2011}. In experiments, the ringlike excitations have been observed in the hyperfine states $|F=1, m_f=-1\rangle$ and $|F=2,m_f=+1\rangle$ of $^{87}$Rb under the rotating cylindrically magnetic trap \cite{Mertes-PRL-2007}. In $|\pm1\rangle$ spinor $^{87}$Rb BECs, more ringlike excitation modes have also been yielded under cylindrically magnetic confinement \cite{Scherer-PRL-2010}. In theoretical simulations, a family of 3D gap solitons of the multiple rings has been reproduced, as supported by 1D optical lattices \cite{Mateo-PRA-2010}.

In addition to the intrinsic vorticity, the MRVS has an additional excitation freedom, corresponding to an infinite number of nodes of the wave function in the radial direction. The early linear stability analytical results \cite{Carr-PRL-2006,Carr-PRA-2006} have shown that, despite of the radial excitation, the 2D (or 3D in the sphere shell) MRVS can be stable in the harmonic confinement below a threshold of the interaction strength. For large interaction strength, the MRVS states are unstable to collapse. When the instability times is much larger than the time scale $2\pi/\omega$, $\omega$ is the harmonic trap frequencies, it is said to be experimentally stable for small interaction strength \cite{Carr-PRL-2006,Carr-PRA-2006}. However, in a recent 3D study, it is found that the MRVS with $n$=2 is unstable against quadrupole perturbations \cite{Mateo-PRA-2010}. This implies that the radial excitation of MRVS could display more rich stability properties if subjected to the 3D geometry. This motivates us to investigate the stability of the MRVS in the 3D space. Besides, considering the cylindrical magnetic confinement in the related experiments \cite{Mertes-PRL-2007, Scherer-PRL-2010}, the full 3D equation is necessary to study the ring vortex solitons.

In this paper, the 3D ring vortex solitons are constructed in the framework of GPE at the cylindrical coordinate for both attractive and repulsive BECs. The stabilities of SRVS and MRVS depend on $S$ differently. The SRVS with $S$=1 has better stability, i.e., large $g_c$, a threshold of the atomic interaction strength below which the ring vortex state is robustly stable against random perturbation, while the SRVS with $S\geq2$ corresponds to the greatly decreased $g_c$ and therefore poor stability. On the contrary, the MRVS with large $S$ has better stability especially when the radial excitation is high. No or very small stable regime is found for the MRVS with $S$=1. The differences between the ground and radially excited states of the ring vortex soliton are further demonstrated in the dynamically unstable evolution. The SRVS shows the simultaneous symmetrical splitting in the transverse plane, while the MRVS shows the periodical expand-merge cycles in the longitudinal direction before collapse. Such dynamical instability undergoes at the timescale of about several seconds, being longer than $2\pi/\omega$, proving their good experimental stability.

The paper is outlined as follows. We first give a family of ring vortex solitions by using Newton continuation method in Sec. II. Next, the stability properties of the ring vortex solitons is analyzed by using linear stability analysis in Sec. III, which is followed by the direct numerical simulations of the perturbed ring vortex solitons in Sec. IV. Finally, we conclude the main results of the work in Sec. V.

\section{Theoretical model and stationary solutions}

We consider the BECs in an external harmonic trapping potential
$V(r,z)$=$m(\omega_r^2 r^2+ \omega_z^2 z^2)/2$, where
$r^2$=$x^2+y^2$, $m$ is the atom mass, $\omega_{r}$ and $\omega_{z}$
are the radial and axial trapping frequencies. The wave function
$\psi$ of the BECs satisfies the dimensionless GPE
\begin{eqnarray}\label{GP1}
i\frac{\partial \psi}{\partial t}&=&-\frac{1}{2}[{\frac {1 }{r}}\frac{\partial }{\partial
r}(r\frac{\partial}{\partial r})+{\frac {1}{{r} ^{2}}}\frac{\partial^2 }{\partial
\theta^2}+\frac{\partial^2}{\partial z^2}]\psi \\ \nonumber
&+&V(r,z)\psi+g \left| \psi \right|  ^{2}\psi+i\Omega\,\frac{\partial \psi}{\partial \theta},
\end{eqnarray}
where $\theta$ is the azimuthal angle, $g$=$4\pi Na_s/a_0$ is the
interaction strength, as determined by the total number of particles
$N$ in the condensate, the $s$-scattering wave length $a_s$, and the
harmonic oscillator length $a_0$=$\sqrt{\hbar/\omega m}$, $\Omega$
is the rotation angular frequency. The dimensionless
$V(r,z)$=$\frac{1}{2}(\gamma_r^2 r^2+\gamma^2_z z^2)$ with
$\gamma_r$=$\frac{\omega_r}{\omega}$,
$\gamma_z$=$\frac{\omega_z}{\omega}$, and
$\omega$=$\min\{\omega_r,\omega_z\}$. Eq.(1) is obtained by
rescaling the length by $a_0$, the time by $\omega^{-1}$, and the
energy by $\hbar \omega$.

Assume the wave function is
\begin{equation}
\psi(r,z,\theta,t)=\phi(r,z)e^{iS\theta-i\mu t},
\end{equation}
where $S$ is azimuthal quantum number, i.e., the intrinsic vorticity and $\mu$ is the chemical potential. The function $\phi(r,z)$ satisfies the equation
\begin{equation}
-\frac{1}{2}[\frac{1}{r}\frac{\partial}{\partial r}(r\frac{\partial}{\partial r})+\frac{\partial^2}{\partial z^2}-\frac{S^2}{r^2}]\phi+V(r,z)\phi-S\Omega+g\phi^3=\mu\phi,
\end{equation}
with the boundary conditions of $\lim\limits_{r\rightarrow 0 \atop |z|\rightarrow 0} \phi(r,z)=0$ and $\lim\limits_{r\rightarrow \infty \atop |z|\rightarrow \infty} \phi(r,z)=0$.
The linear-limit solution of Eq. (3) is written as the scaled linear combinations of products of the wave functions of the harmonic oscillator \cite{Bao-SIAM-2009},
\begin{equation}
\phi(r,z)=\frac{\gamma_r^{(S+1)/2}\gamma_z^{1/4}}{\sqrt{\pi
C^{S}_{n}}\pi^{1/4}} r^{S}\hat{L}_{n}^{S}(\gamma_r
r^2)e^{-\frac{1}{2}(\gamma_r r^2+\gamma_z z^2)},
\end{equation}
where $C^{S}_{n}=\prod_{j=1}^{S}(n+j)$, $\hat{L}_{n}^{S}$ is the generalized-Laguerre polynomials, and $n$ is the quantum number in the $r$ direction (the quantum number in the $z$ direction has set to be zero). The chemical potential corresponding to Eq. (3) and Eq. (4) is written as
\begin{equation}
\mu_{nS}=\gamma_r[2(n-1)+S+1]-S\Omega+\frac{1}{2}\gamma_z.
\end{equation}
The stationary soliton solutions can be obtained as numerical solutions of Eq. (3) using the Newton continuation method \cite{Bao-JCP-2003} with $\mu_{nS}$ input, starting with the linear-limit solutions, under the constraint that the normalization of the wave function $\tilde{N}$ and the energy $E$ are conserved with numerical iteration time, where
\begin{equation}
\tilde{N}(\psi)=\int_{\textbf{R}}|\psi(\textbf{R},t)|^2 d\textbf{R}=\tilde{N}(\psi_{0})=1, t\geq0,
\end{equation}
and the energy
\begin{eqnarray}
E_{g,\Omega}(\psi)&=&\int_{\textbf{R}}[\frac{1}{2}|\psi(\textbf{R},t)|^2+V(r,z)|\psi(\textbf{R},t)|^2 \\ \nonumber
&+&\frac{g}{2}|\psi(\textbf{R},t)|^4-\Omega S] d\textbf{R}\equiv E_{g,\Omega}(\psi_0), t\geq0.
\end{eqnarray}

\begin{figure}
\begin{center}
\includegraphics[width=\linewidth]{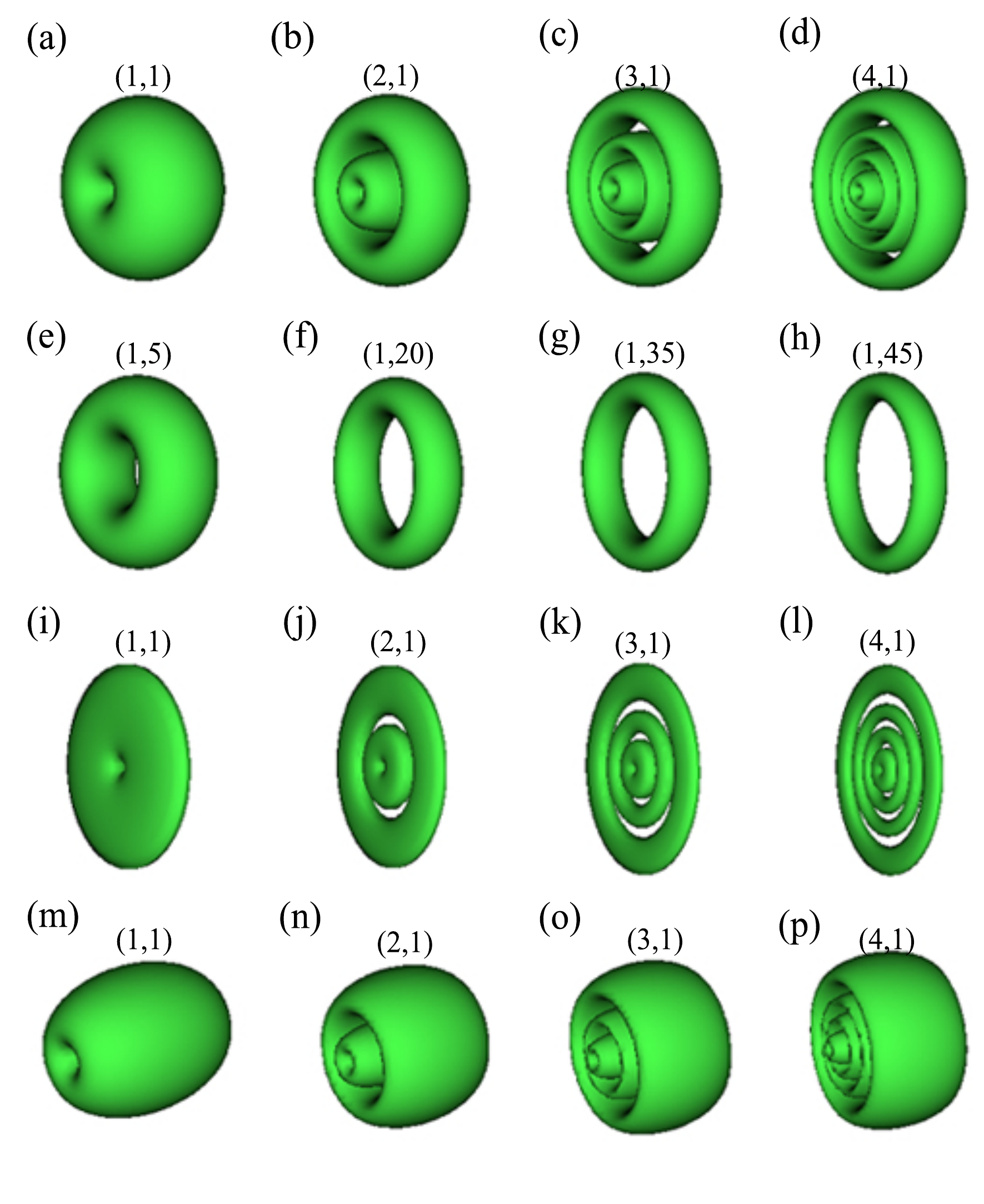}
\caption{The 3D ring vortex solitons with different quantum-number
sets ($n,S$) found numerically for attractive BEC, as visualized by
isosurface of dimensionless density of the wave function in the
isotropic (a)-(h), pancake (i)-(l), and cigar (m)-(p) harmonic
traps, respectively.}
\end{center}
\end{figure}

Shown in Fig. 1 are various ring vortex solitons obtained in the case of the attractive interaction under different shaped harmonic potentials. For the isotropic harmonic potential of $\gamma_r$=$\gamma_z$=1, $g$=-0.0083$N$ refers to a $^7$Li system both with $\omega_r$=$\omega_z$=$20\pi$Hz, for the cigar-shaped harmonic potential of $\gamma_r$=1.44 and $\gamma_z$=1, $g$=-0.0117$N$ refers to $^7$Li with $\omega_r$=178Hz and $\omega_z$=123Hz \cite{Bradley-PRL-1995}, and for the pancaked harmonic potential of $\gamma_r$=1 and $\gamma_z$=20, $g$=-0.0083$N$ refers to $^{7}$Li with $\omega_r$=20$\pi$Hz and $\omega_z$=400$\pi$Hz \cite{Rychtank-PRL-2004}. In the case of isotropic harmonic potential, we also study the ring vortex soltons for the repulsive interactions, where $g$=0.0188$N$ refers to $^{87}$Rb BECs with $\omega_r$=$\omega_z$=$20\pi$Hz. In this work, $\Omega$=0.7 is fixed. The vortex solitons obtained for the repulsive interactions (not shown here for the sake of brevity) resemble those shown in Fig. 1 for the attractive interaction.

The ring vortex solitons are further illustrated according to the radial profile of the wave function. Fig. 2(a) shows that the influence of the quantum number $n$ on the ring vortex solitons, i.e., the radially excited level is increasing with $n$, corresponding to more nodes of the wave function in the radial direction. Fig. 2(b) shows that the influence of the quantum number $S$ on the ring vortex solitons. The ring vortex solitons with the larger $S$ demonstrates the deeper depletion of density at core area (denoted by dash arrows) and the lower density peak (denoted by solid arrows). Fig. 2(c) shows the influence of the sign of the interaction nonlinearity on the shape of the ring vortex solitons for a given interaction strength. The density of the wave function at the first peak (denoted by the arrow in Fig. 2(c)) in the case of the attractive interaction is higher a little than that in the case of the repulsive interaction. Such difference is the most prominent in the case of ($n,S$)=(1,1), but is weak in the cases of large $n$ or large $S$. Fig. 2(d) illustrates variation of the radial profile of the ring vortex solitons under the different shaped harmonic potentials.

\begin{figure} 
\begin{center}
\includegraphics[width=\linewidth]{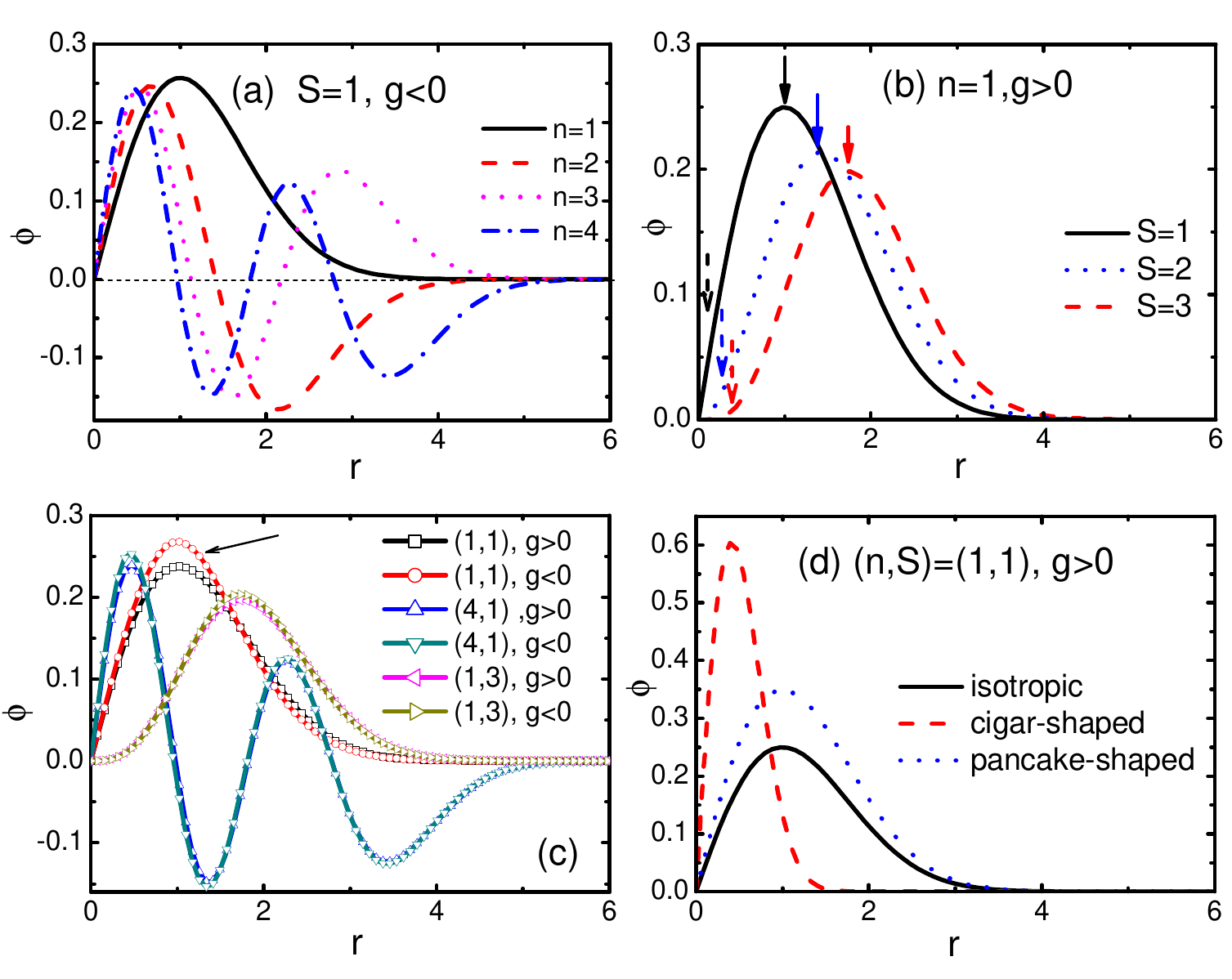}
\caption{ The radial density profile of the ring vortex solitons
characterized by the quantum numbers ($n,S$) for the attractive
($g<0$) or repulsive ($g>0$) interactions under the isotropic
harmonic potential in (a)-(c) and also the cigar-shaped and
pancake-shaped harmonic potentials in (d).}
\end{center}
\end{figure}

Shown in Fig. 3 is the $\mu(N)$ curves of the ring vortex solitons. For a given $N$, as compared with the SRVS family with $n$=1, the MRVS families with $n$=2, 3, 4, $\dots$ are of increasing $\mu$. Within a SRVS or MRVS family, $\mu$ is increasing with $S$. Each data shown in Fig. 3 corresponds to a convergent stationary state, which is further investigated for the stability and dynamical evolution.

\begin{figure} 
\begin{center}
\includegraphics[width=\linewidth]{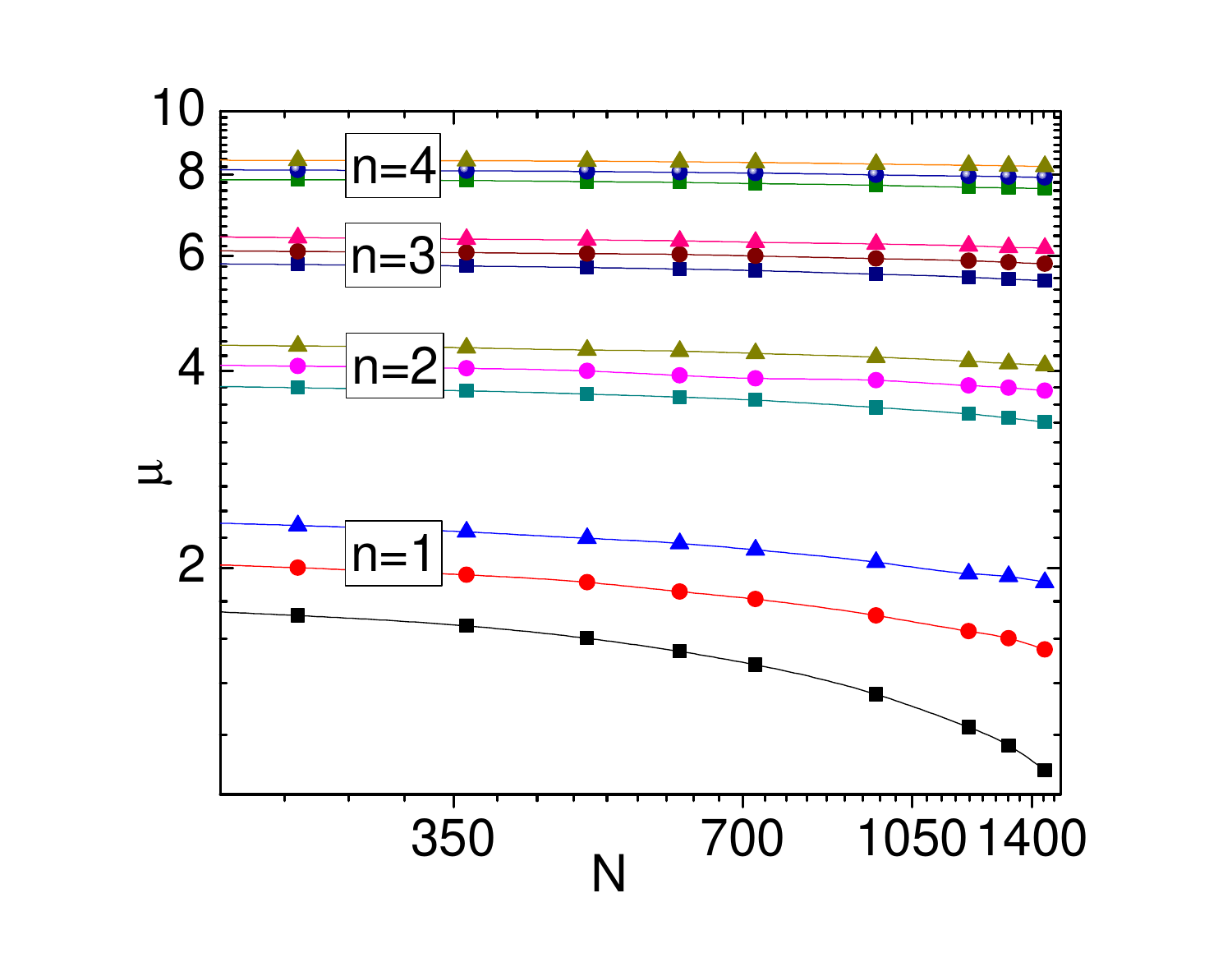}
\caption{(Color online) The $\mu(N)$ curves of the ring vortex
solitons of $n$=1 ,2, 3, 4 with $S$=1 (square), $S$=2 (circle), and
$S$=3 (triangular), as calculated with $\gamma_r$=$\gamma_z$=1 and
$g$=-0.0083$N$, and $\Omega$=0.7, referring $^7$Li condensate under
the isotropic harmonic potential with
$\omega_r$=$\omega_z$=20$\pi$Hz.}
\end{center}
\end{figure}

\section{stability analysis}
The stability of the ring vortex solitons is analyzed using the linear stability analysis. The wave function that deviates slightly from the stationary solutions is constructed as
\begin{equation}
\psi=[\phi(r,z)+ue^{iE t}+w^{*}e^{-iE^{*}t}]e^{iS\theta-i\mu t},
\end{equation}
where $|u|$, $|w|\ll1$ are eigenmodes. Substituting Eq. (8) into Eq. (1) gives
\begin{equation}
\left( \begin{array}{cc}L & -g\phi^2\\g\phi^2&-L\end{array}\right)\left(\begin{array}{c}u\\w\end{array}\right)=E\left(\begin{array}{c}u\\w\end{array}\right),
\end{equation}
where
$L\equiv(\partial_{rr}+\frac{1}{r}\partial_{r}+\partial_{zz}-\frac{S^2}{r^2})/2-V(r,z)-2g\phi^2+S\Omega+\mu$, $E$ is the eigenvalue related to $u$ and $w$, which is obtained by
diagonalization of Eq. (9) under the boundary conditions demanding that $u(r,z),w(r,z)\rightarrow0$ at $r,|z|\rightarrow\infty$ and $r,|z|\rightarrow0$. Applying Eq. (9), we analyze the ring vortex solitons of different radial excited states, not only the SRVS, i.e., $n$=1, but also the MRVS, from the first radially excited state, i.e., $n$=2, to the higher radially excited states.

\begin{figure}
\includegraphics[width=\linewidth]{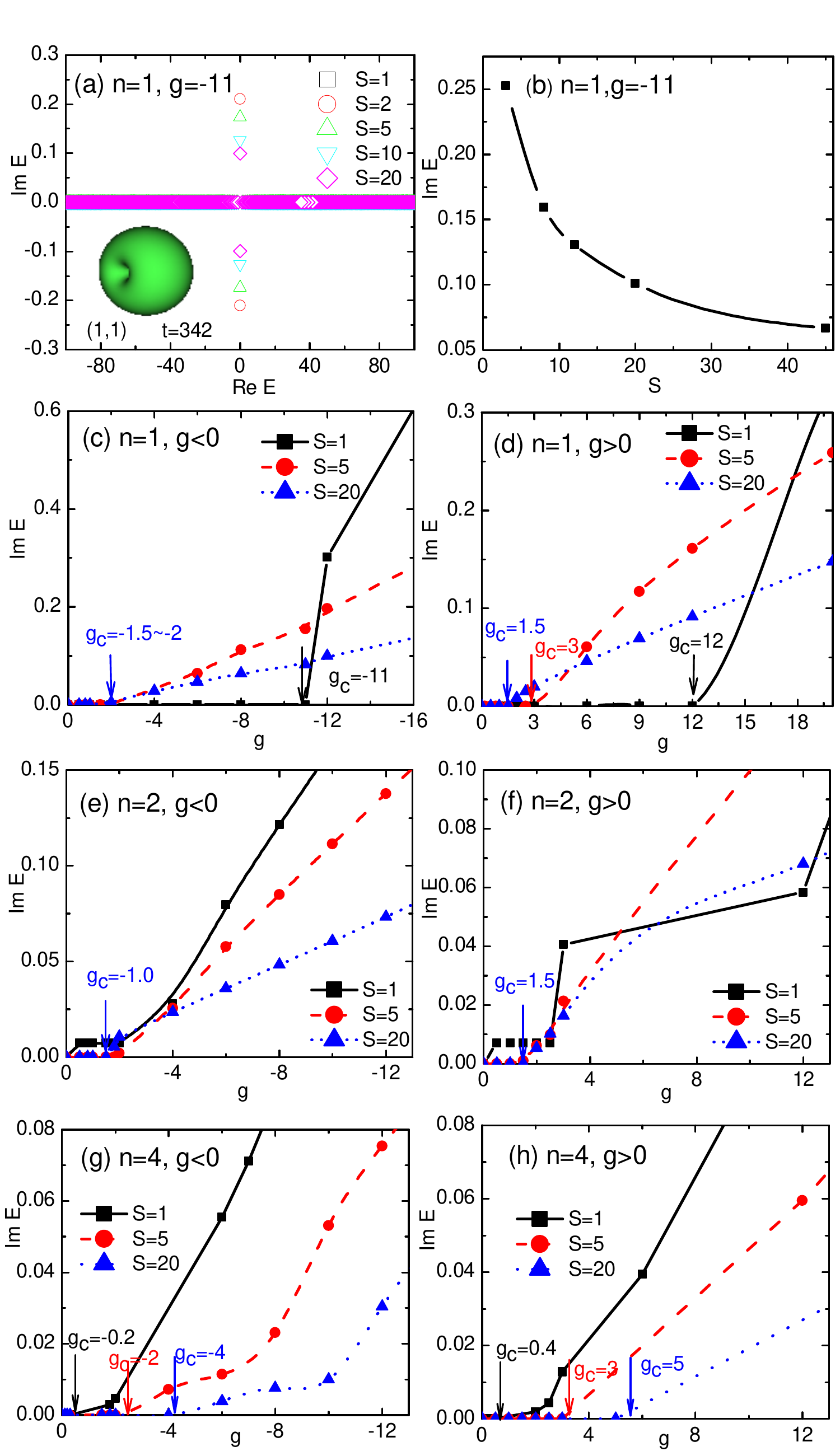}
\caption{(Color online) The imaginary part in energy spectrums (Im
E) of the stability analysis, (a)-(b): Im E .vs. $S$ obtained for
SRVS under a given attractive interaction $g$=-11, inset in (a)
displays a robust SRVS wit $S$=1 at dimensionless time $t$=342
against noise perturbation, and Im E .vs. $g$ obtained for SRVS
[(c)-(d)] and MRVS of $n$=2 [(e) and (h)] and $n$=4 [(g) and (h)]
with different $S$. }
\end{figure}

\begin{figure} 
\begin{center}
\includegraphics[width=\linewidth]{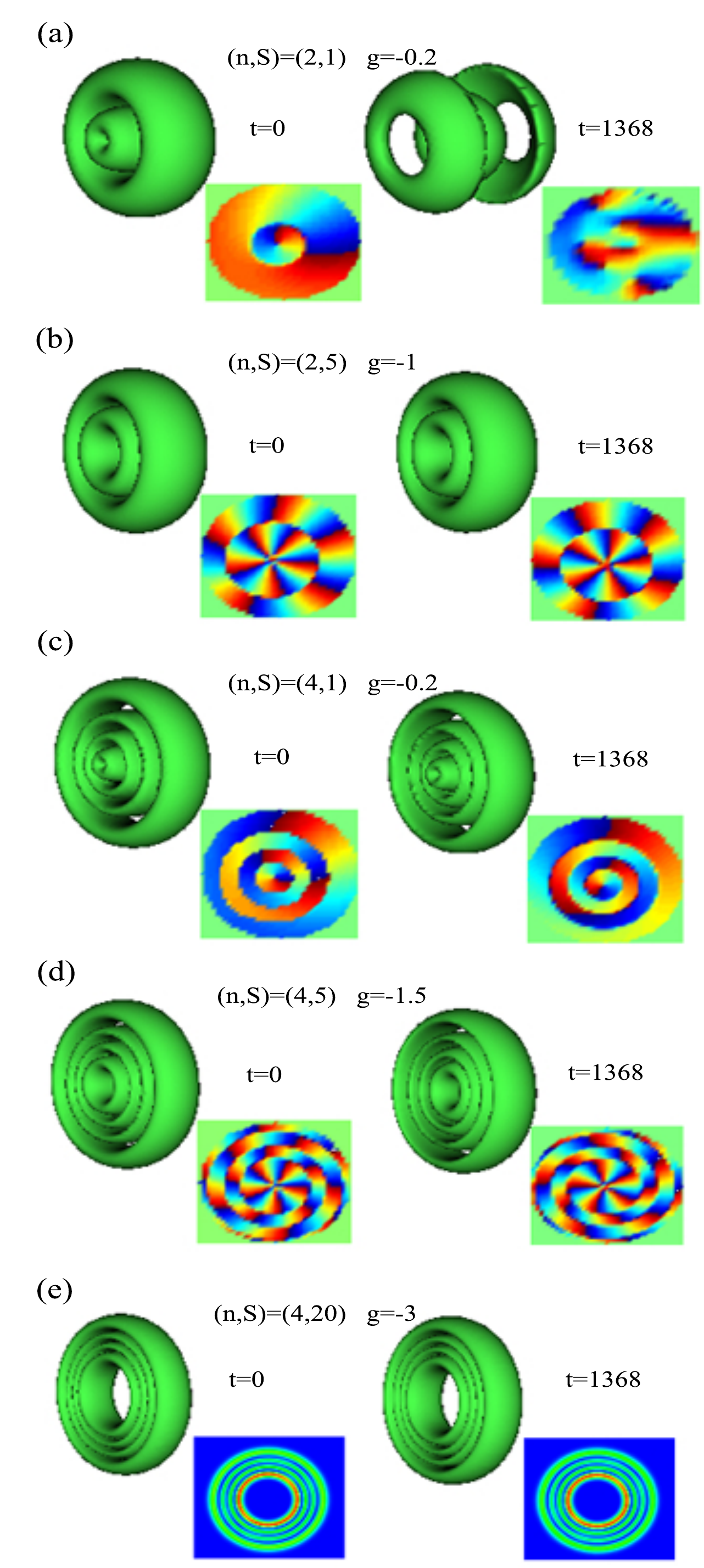}
\caption{(Color online) The dynamical evolution of the MRVS in
unstable and stable regions in Fig. 4 (e) and (f), (a) is
dynamically unstable and (b)-(e) are dynamically stable states,
which are consistent with linear stability analysis. Referring a
$^{7}$Li system confined by $\omega_{r}$=$\omega_{z}$=20$\pi$ Hz,
the size is about 120$\mu$m and the real time $t'$=$0.0159t$ second,
$t$=0 corresponds to the initial state of a 1\% random perturbation
in density.}
\end{center}
\end{figure}

Fig. 4(a) shows the real and imaginary parts of eigenvalues of the SRVS obtained for a given interaction strength in an isotropic harmonic potential. When $S$=1, all eigenvalues are real, indicating the corresponding state is stable. When $S\geq$2, complex eigenvalues emerge, indicating the corresponding states are dynamically unstable. The inverse of the imaginary part of the complex eigenvalue, Im E, gives the time scale of such a dynamical instability. As shown in Figs. 4 (a) and 4(b), it is found that Im E decreases with $S$, which indicates that the ring vortex soliton with the larger intrinsic vorticity is of the longer lifetime before collapse, i.e., the better experimental stability.

Moreover, Figs. 4(c) and 4(d) show the $g$ dependence of Im E obtained for the SRVS with different $S$. The eigenvalues become complex above the critical interaction strength $g_c$, showing the onset of the dynamical instability. The $g_c$ value is dependent on $S$. For example, our results show that $g_c$ is around -11 when $S$=1 but decreases to -1.5 and -2 when $S$=5 and 20 in the case of the attractive interaction. The corresponding value of $g_c$ becomes a little larger in the case of the repulsive interaction, being around 12, 3, 1.5 for the SRVS with $S$=1, 5, and 20, respectively. The case of the SRVS, i.e., the general ring vortex soliton without radial excitation, has been studied previously \cite{Pu-PRA-1999,Ueda-PRL-2002,Mihalache-PRA-2006}. It was found that for sufficiently weak interaction strength the SRVS with $S$=1 is stable, while for $S\geq$2 the SRVS is unstable to quadrupole oscillations. For the instability times being much longer than the time scale $2\pi/\omega$ of the BECs, the SRVS with $S\geq$2 is said to be experimentally stable for small interaction strength. Our results shown in Figs. 4(a)-(c) are consistent with the previous study for the general ring soliton without the radial excitation \cite{Pu-PRA-1999,Ueda-PRL-2002,Mihalache-PRA-2006}.

Figs. 4(e) and 4(f) show the curve of Im E .vs. $g$ obtained for the MRVS of $n$=2, i.e., the first excited state. The stability of the ring vortex soliton of the first radial excited state has been studied based on the 2D solutions \cite{Carr-PRL-2006}, which shows that the first radial excited state with $S$=1 is stable for the sufficiently small interaction strength below a critical value. In our 3D study, for $S$=1, no stable regime is found for the first excited state. This result means that the 3D MRVS of the first excited state with $S$=1 could be different from its counterpart in 2D. The critical dimensionality for the GPE is 2D. Further, we find that, for large $S$, the 3D MRVS can be made stable. For example, we gives two Im E .vs. $g$ curves obtained for MRVS with $S$=5, and $S$=20 in Figs. 4(e) and (f), where the stable regime, though being very small, is found up to $g_c$=-1.0 in the cases of the attractive interaction and $g_c$=1.5 in the case of the repulsive interactions. Such results indicate that the stability of the MRVS is dependent on the intrinsic vorticity. The MRVS with large $S$ is expected to stable.

Shown in Figs. 4(g) and 4(h) are results for the MRVS of $n$=4. The $S$ dependence of the stability becomes more prominent when the MRVS is of the higher radial excitation. The stable region, though being very small, below $g_c$=-0.2 and $g_c$=0.4, are found for the $n$=4 MRVS with $S$=1 under the attractive and repulsive interactions. The stable region increases for large $S$. When $S$=5, the stable regions are up to $g_c$=-2 and $g_c$=3 in the cases of the attractive and repulsive interactions respectively. Further, when $S$=20, the stable regions increase up to $g_c$=-4 and $g_c$=5 in the cases of the attractive and repulsive interactions respectively. Through comparing two cases of $n$=2 [Figs. 4(e)- 4(f)] and $n$=4 [Figs. 4(g)-4(h)], we can find that the intrinsic vorticity tends to improve the stability of the MRVS, and such effect turns more significant when the radial excitation of MRVS is high. The MRVS of large $S$ has the prominent stable region.

The linear stability analytic results for the $S$ dependence of the MRVS stability is further confirmed by the directed simulation based on Eq. (1). Corresponding to Fig. 4(e), Fig. 5(a) and (b) proves that the radially first excited state of MRVS of $n$=2 is unstable against random perturbation when $S$=1, but turns to stable when $S$ is large, such as $S$=5, under a small $g$. Next, corresponding to Fig. 4(g), Fig. 5(c)-(e) prove the stable MRVS of $n$=4 with different $S$ under the given interaction strength below the increasingly large $g_c$.

\section{dynamical evolution}
The ring vortex solitons are unstable against the dynamical instability when the interaction strength is larger than $g_c$. Such dynamical instability is simulated through numerically integrating
Eq.(1) using the time-splitting-spectral technique \cite{Bao-JCP-2006}. In our simulation, the stationary solutions of the ring vortex solitons are used as initial states after adding a random
perturbation of a 1\% relative amplitude of density.

\begin{figure}
\centerline{\includegraphics[width=\linewidth]{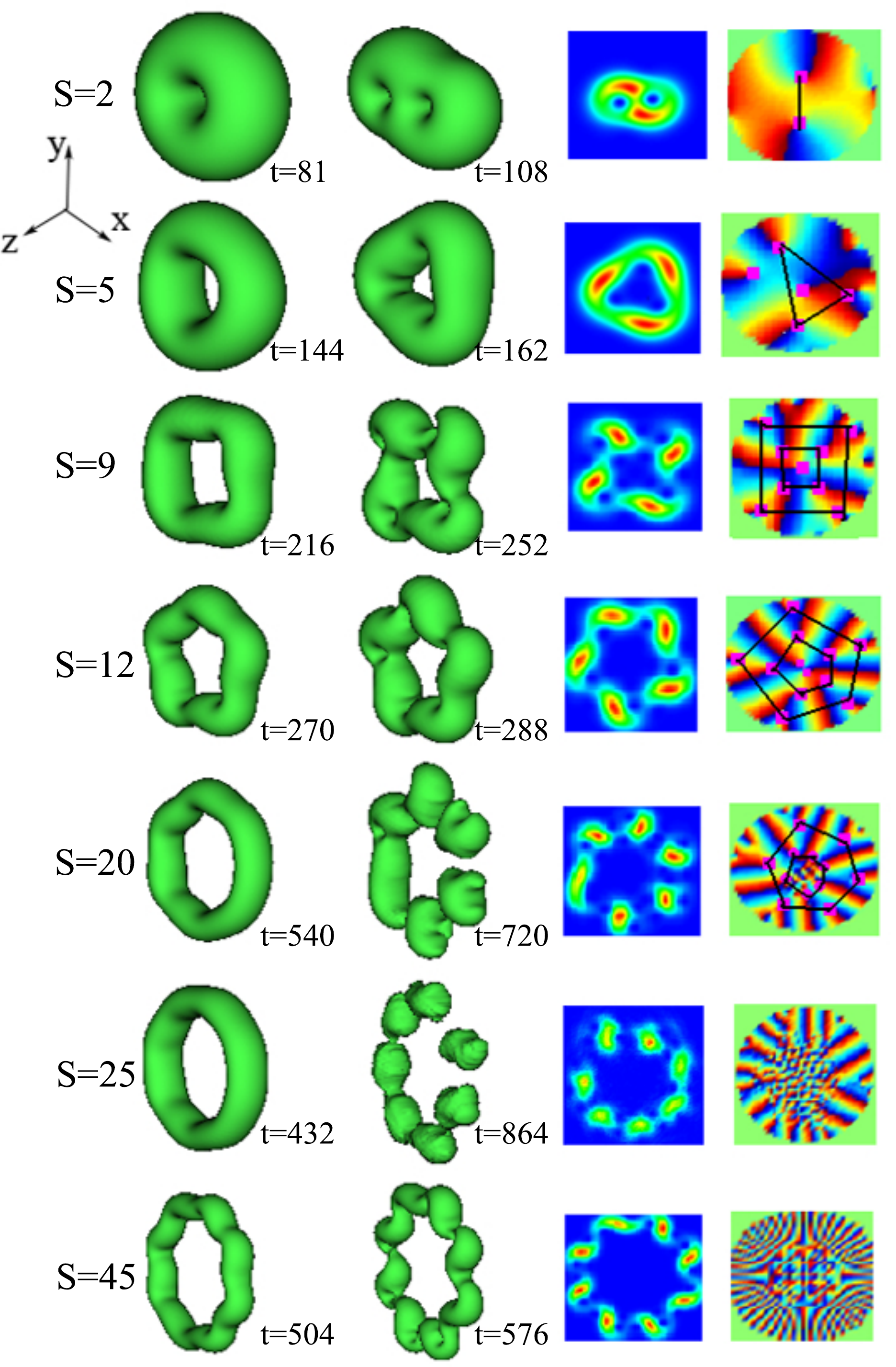}}
\caption{(Color online) Noise-induced splitting of the ring vortex
solitons of different $S$ at the dimensionless time $t$, and the
corresponding density and phase (the third and forth columns) plots
in the $x$-$y$ plane. To translate the results into the
experiment-related units, we assume a $^{7}$Li condensate containing
about 1500 atoms in a isotropic harmonic potential with
$\omega_{r}$=$\omega_{z}$=20$\pi$ Hz. The radius of the ring vortex
solitons here are about 120$\mu$m, decaying at the real time
$t'$=$0.0159t$ second.}
\end{figure}


\begin{figure}
\begin{center}
\includegraphics[width=\linewidth]{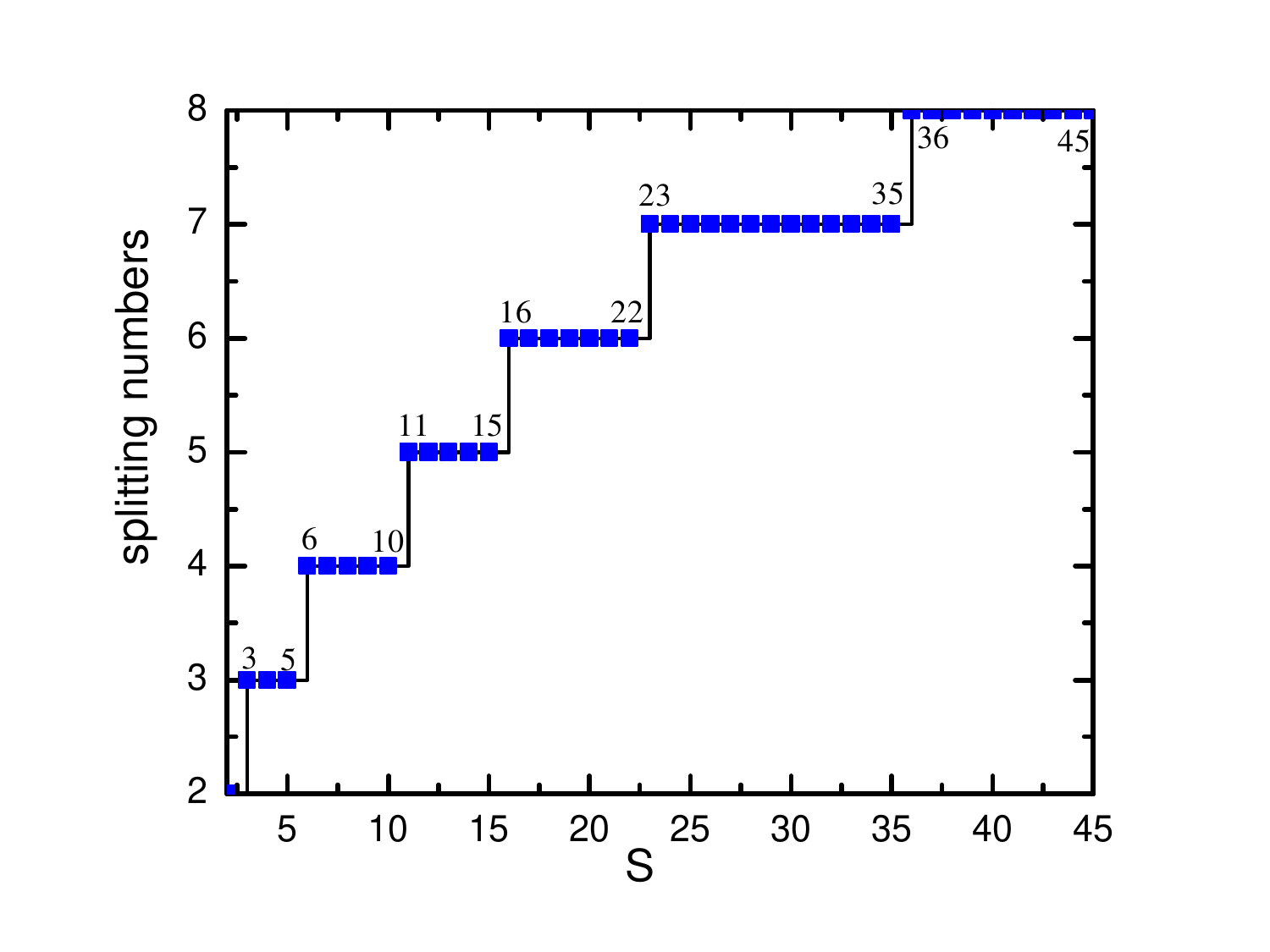}
\caption{(Color online) Noise-induced splitting of giant vortices.
The splitting numbers as a function of different $S$ for the
attractive interaction are shown.}
\end{center}
\end{figure}

We first simulate the dynamical evolution of the SRVS. As shown in Fig. 6, the SRVS evolves into polygonal rings, then splits into small fragments, and finally collapses. The multiple symmetry is revealed during such evolution. As $S$ increases, we obtain the 2-, 3-, 4-, 5-, 6-, 7-, and 8-fold symmetry, as shown in Figs. 6 and 7. As seen in the phase profile of the wave function, the multiply quantized vorticity splits into many singly quantized vortices. Such singly quantized vortex cannot be seen in the density plot, and hence they are called as hidden (or ghost) vortices \cite{Makoto-PRA-2002,Wen-PRA-2010}. The singly quantized vortices are self-organized into the regular lattice, contributing to the symmetry in splitting. Such symmetry in instability can be regarded as simultaneous one, because the random initial perturbation we add is of no azimuthal mode. The multiple symmetrical splitting phenomena have been demonstrated for giant vortices
\cite{Kawaguchi-PRA-2004,Mottonen-PRL-2007,Isoshima-PRL-2007,Kuopanportti-PRA-2010}. The symmetry in splitting and the number of the fragments are considered to be equal to the added angular-momentum quantum number of the Bogoliubov excitation mode responsible for the splitting \cite{Kawaguchi-PRA-2004,Kuopanportti-PRA-2010}. Ones find a total of three types of splitting patterns, 2-fold, 3-fold, and 4-fold symmetries, when the initial perturbation is random noise in the density of the wave function \cite{Isoshima-PRL-2007,Kuopanportti-PRA-2010}. Figs. 6 and 7 indicate that the symmetry in splitting of the ring vortex soliton can have more types. As matter of fact, the number of symmetry can be manipulated by the interaction strength $g$. As shown in Fig. 8, the number in symmetry (or the number of the fragments) decreases with $g$ in the case of the repulsive interaction and increases with $g$ in the case of the attractive interaction.

\begin{figure} 
\begin{center}
\includegraphics[width=\linewidth]{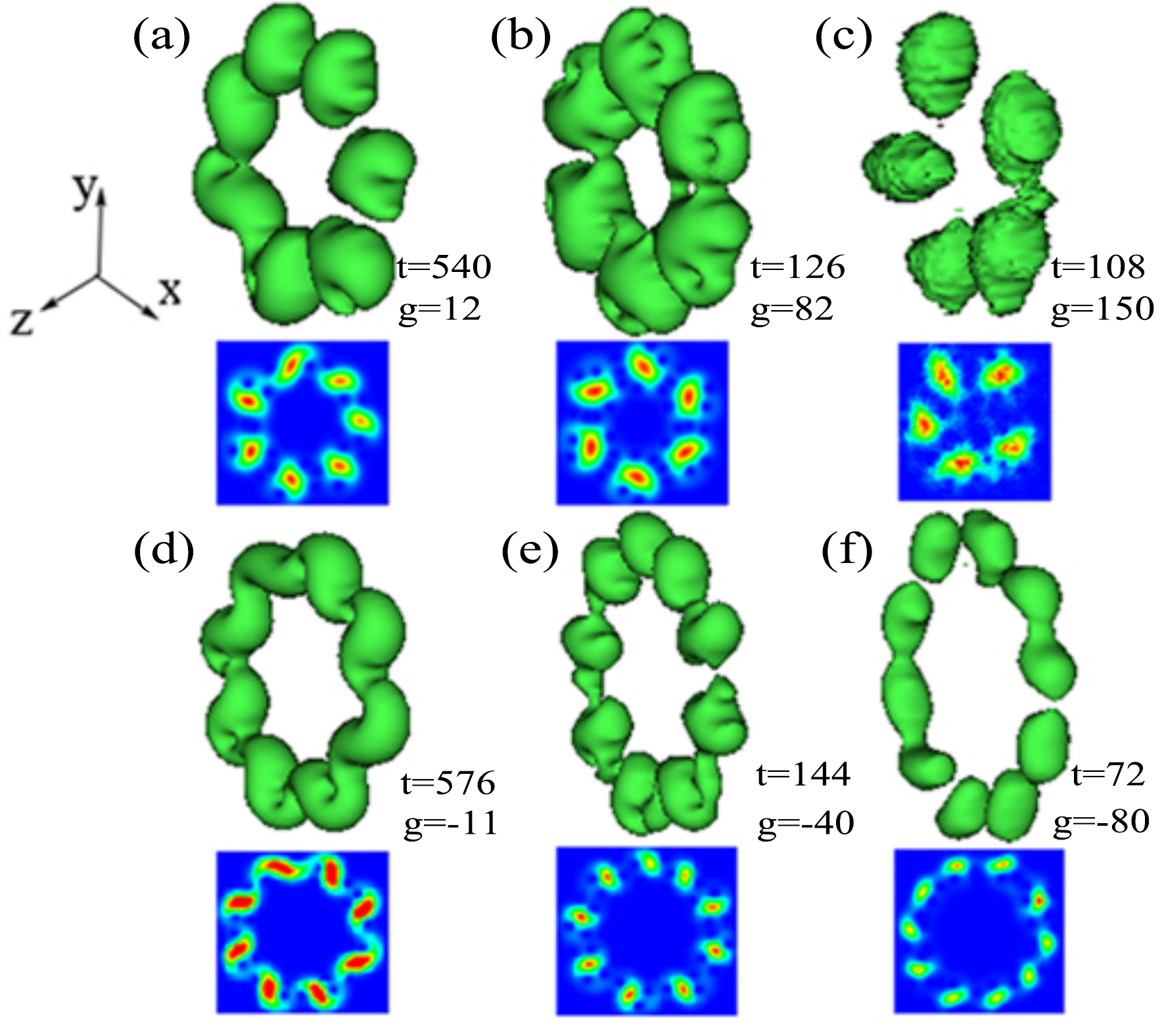}
\caption{(Color online) Noise-induced splitting of the 3D vortex
solitons of large vorticity in (a)-(c) with $S=25$ for the repulsive
interaction and (d)-(f) with $S$=45 for the attractive interaction.
To translate the results into the experiment-related units, we
assume the $^{87}$Rb condensate containing about 640 (a), 4360 (b),
and 8000 (c) atoms and the $^7$Li condensate containing 1325 (d),
4820 (e), and 9638 (f) atoms in a isotropic harmonic potential with
$\omega_{r}$=$\omega_{z}$=20$\pi$ Hz. The radius of the ring vortex
solitons here are about 31$\mu$m for (a)-(c), and 120$\mu$m, being
snapshots at the real time $t'=0.0159t$ second.}
\end{center}
\end{figure}

\begin{figure} 
\begin{center}
\includegraphics[width=\linewidth]{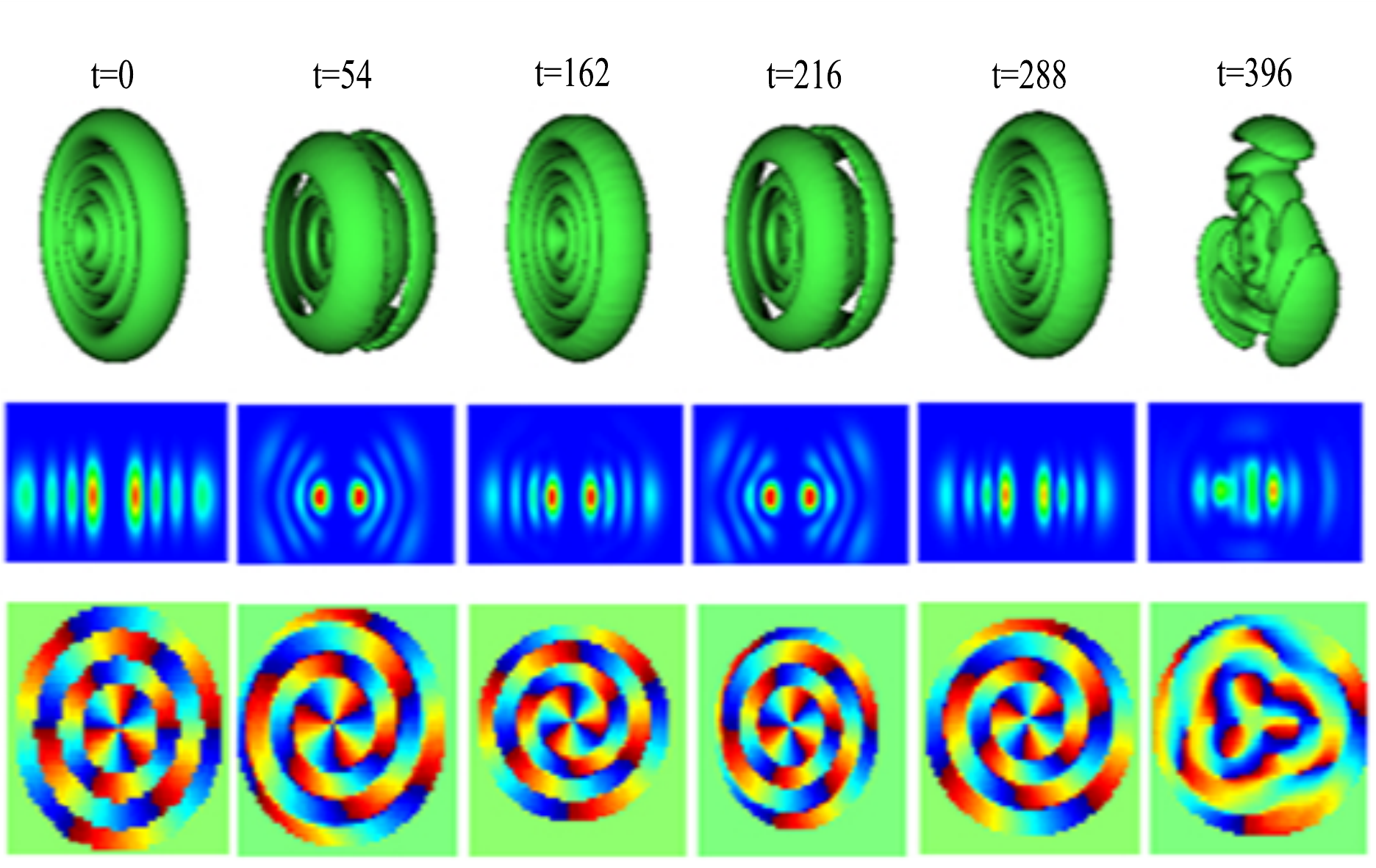}
\caption{(Color online) Noise-induced splitting of a vortex soliton
of (n,S)=(4,3) at the dimensionless time $t$ and the corresponding
density plot in the $x$-$z$ plane (middle row) and phase plot in the
$x$-$y$ plane (lower row). To translate the results into the
experiment-related units, we assume a $^{7}$Li condensate containing
about 1500 atoms in a isotropic harmonic potential with
$\omega_{r}$=$\omega_{z}$=20$\pi$ Hz. The radius of the ring vortex
solitons here are about 83.68$\mu$m, decaying at the real time
$t'$=$0.0159t$ second.}
\end{center}
\end{figure}

Second, we simulate the dynamical evolution of the MRVS. As shown in Fig. 9, the MRVS demonstrates expand-merge cycles, then collapses. The similar phenomenon, expand-shrink cycles occurring in the
transverse direction, was predicted for the dark SRVS \cite{Ueda-PRL-2002,Mihalache-PRA-2006,Salasnich-PRA-2009}. As compared with the dark SRVS, a principal difference is that the expand-merge cycles of the 3D MRVS occurs in the longitudinal direction. The condensate expands along the axial direction. The outer rings expand faster than the internal rings. At $t$=54, the outmost ring splits into two pieces, and the internal rings expand but remain united. In contrast, the inmost ring shrinks, as seen in the density profile in the longitudinal plane shown in Fig. 9, middle row. The expanded condensates and pieces subsequently unite to restore the original shape at $t$=162. Such cycles repeats two times, then collapse finally. The expand-merge cycles reflect the expansion in the axial direction and the oscillation in the radial direction of the density of the wave function. Besides, during the dynamical evolution of the MRVS, as seen in the phase profile of the wave function in the transverse plane shown in Fig. 9, lower row, the multiply quantized vorticity is maintained well during the expand-shrink cycles. No vortex splitting is observed until the final collapse, which indicates that for the MRVS, the instability of the multiple radial waves prevails over the azimuthal instability. The similar expand-merge evolutions are demonstrated in the case of the repulsive interaction, where the cycles repeat three times before collapse, indicating the longer lifetime.

We also simulate the dynamical instability under the pancake-shaped and cigar-shaped harmonic potentials. The expand-merge cycle occurring for the isotropic harmonic potential is suppressed under the pancake confinement potential. Under the cigar-shaped harmonic potential, being elongated in the axial direction, the MRVS does not demonstrate expand-merge cycles either. Besides, another interesting dynamics phenomenon, intertwining of vortices in the cigar-shaped BECs \cite{Mottonen-PRA-2003,Mateo-PRL-2006,Huhtamaki-PRL-2006}, is also not observed in our simulations.

\section{discussion and conclusions}
Consider the cylindrical magnetic confinement, we use fully 3D equation to investigated the stationary state, stability, and dynamical evolutions of the ring vortex solitons in the attractive and repulsive BECs. A family of stationary ring vortex solitons are obtained numerically by Newton continuation method. The stability properties of the ring vortex solitons are predicted for a given interaction strength by using the linear stability analysis and confirmed further by the direct simulation. The stabilities of the SRVS and MRVS depend on $S$ differently. The SRVS with $S$=1 corresponds to a large $g_c$, a threshold below which the solution is stable against random perturbation, while the SRVS with $S\geq2$ corresponds to the greatly decreased $g_c$. The $S$ dependence of the stability of the MRVS is on the contrary. The prominent stable regimes are found for the MRVS of $n$=4 with large $S$, such as $S$=5 and 20, while very small stable regimes are obtained for the MRVS of $n$=4 with $S$=1, and even no stable regime is found for MRVS of $n$=2 with $S$=1.

Therefore, we can expect the robust dynamical stability against
random perturbation for the SRVS with $S$=1 and the MRVS with large
$S$, when the atomic interaction strength is less than $g_c$. For
the radial ground state of the ring vortex solition, the value of
$g_c$ are around -11 and 12, as predicted for the SRVS with $S$=1
under the attractive and repulsive interactions. For the radially
excited state, the values of $g_c$ are around -4 and 5, as predicted
for the MRVS with $S$=20 under the attractive and repulsive
interactions. In our study, we consider the cylindrical magnetic
trap with the harmonic frequency $\omega_r$=$\omega_z$=$20\pi$Hz,
which corresponds to a critical atom number of the stable ring
vortex soliton being about $600 \sim 1300$ in the case of the
radially ground state (where, we refers to the $^{87}$Rb system with
$a_s$=51${\AA}$ and $m$=1.44$\times10^{-25}$kg and the $^7$Li system
with $a_s$=-79.35${\AA}$ and $m$=1.1702$\times10^{-26}$kg,
respectively) and several hundreds in the case of the radially
excited state. To decrease the magnetic harmonic frequency down to
several Hz, one can expect that the critical atom number increases
to $10^3\sim10^4$. Our results suggest the possible conditions that
the ring vortex soliton can be make stable against random
perturbation.

On the other hand, our study also suggest some most unstable states of the ring vortex solitons, for example, the SRVS with $S\geq2$ and the MRVS with $S$=1. Our direct simulations show that the lifetime of the dynamical evolution can be several seconds before the final collapse, for example the SRVS shown in Fig. 6 and the MRVS shown in Fig. 9 contain about $10^3$ atoms can survive up to 1.5-8 seconds if referring to a magnetic trapping potential with frequency $\omega$=$20\pi$. Such timescales are far larger than the time scale $2\pi/\omega$ \cite{Carr-PRL-2006,Carr-PRA-2006}. Therefore, the dynamical evolution results further suggest that the ring vortex solitons could be observed in experiments if subjected an appropriate interaction strength. The differences between the ground and radially excited states of the ring vortex soliton are also reflected in the different evolution ways before collapse. In the dynamically unstable regime subjected to the large atomic interaction, the SRVS demonstrates the simultaneous symmetrical splitting in the transverse plane, while the MRVS exhibits the periodical expand-merge cycles in the longitudinal direction.

Our results suggest possibilities for creation and observation of robust 3D ring vertex solitons in the cylindrical geometry under the magnetic harmonic confinement. The present work is conducted under the rotating GPE frame, which facilitates the ongoing study that considers the possible dependence of the stability and dynamical evolution of the ring vortex soliton on the rotation angular frequency.

\section{Acknowledgements}
This work was supported by NKBRSF of China (Grant No. 2012CB821305, 2011CB921502) and NSFC (Grant No. 10974228, 11001263).

\end{document}